\documentclass[12pt,a4paper]{revtex4}
\usepackage{epsfig}                           
\begin{document}
\title{Supersymmetry and Fokker-Planck dynamics in periodic potentials}
\author{Mamata Sahoo$^{1}$, Mangal C. Mahato$^{2}$ and A. M. Jayannavar$^{1}$}
\address{$^{1}${Institute of Physics, Bhubaneswar-751005, India}}
\address{$^{2}${Department of Physics, North-Eastern Hill University,
 Shillong-793022, India}} 
\begin{abstract}
Abstract:~~ Recently, the Fokker-Planck dynamics of particles in periodic
potentials $\pm V$,~ have been investigated by using the matrix continued
fraction method.~It was found that the two periodic potentials, ~ one 
being bistable and the other metastable give the same diffusion 
coefficient in the overdamped limit.~  We show that this result 
naturally follows from the fact that the considered potentials in 
the corresponding Schr\"{o}dinger equation form supersymmetric 
partners.~  We show that these differing potentials ${\pm}V$ ~ also 
exhibit symmetry in current and diffusion coefficients:
$J_{+}(F)=-J_{-}(-F)$ and ~ $D_{+}(F)=D_{-}(-F)$ in the presence 
of a constant applied force F. ~ Moreover, ~we show numerically
that the transport properties in these potentials are related even 
in the presence of oscillating drive.\\
\end{abstract}
\maketitle

\vspace{3.5cm}
\noindent
Corresponding Author:  A.M. Jayannavar\\
Email address       :  jayan@iopb.res.in 
\newpage
\section{Introduction}
Periodic potentials are common occurrences in diverse kinds of extended  systems 
such as crystals in solids and microtubules in living systems.~ The nature of the potential profile depends on the medium and the particle  moving through it.
A $Ag^{+}$ ion,~ for example,~ of a AgI crystal in the superionic phase[1] moving 
in a certain direction may be gainfully modeled to encounter a periodic potential 
profile $V(x)=A \cos{kx}+B \cos{2kx}$ with two wavevector components instead of a simple cosinusoidal potential.~  Such potential profile modeling could be based 
on the consideration of crystal structure and the underlying energy landscape  in 
the direction of motion of the ion.~ In some other direction,~ however,~ the periodic potential profile could be entirely different.~ The potential profile 
$-V(x)$ is also periodic with the same periodicity as $V(x)$ yet they have important differences.~ The potential $V(x)$ is bistable in nature,~ whereas $-V(x)$ has a metastable well within a period,~~Fig.1.~

Recent studies show that an overdamped Brownian particle exhibits identical diffusive behavior in the potentials $V(x)$ and $-V(x)$ in the entire range of parameter $\frac{A}{B}$ [2].~  However,~  the diffusion constant of underdamped Brownian particles are quite different for the two related potentials.~ In this work we reason and explain that this result follows from the fact that the potentials of the Schr\"{o}dinger equation with (Euclidian time) corresponding to Fokker-Planck equation of motion [3] with potentials $\pm V(x)$ form supersymmetric pairs and hence have identical eigenvalue spectra[4,5,6].~  The identical eigenvalue spectra along with one to one correspondence between their eigenfunctions ensure that both the systems~~(i.e.,~~with $\pm V(x)$) exhibit same long time average asymptotic behavior.~ This also implies that when a constant force F is applied to the potentials $\pm V(x)$ the average stationary current will satisfy $J_{+}(F)=-J_{-}(-F)$  and  the diffusion constant 
$D_{+}(F)=D_{-}(-F)$.~ This result can also be seen  from the analytic solution of the Fokker-Planck equation.~In addition,~ we show that when the system is driven periodically,~  both the potentials show similar diffusive behavior.
\section{The Model}
         Consider the motion of a particle moving in a tilted periodic potential
$V(x)$ and described by the overdamped Langevin equation[3]
\begin{equation}
\gamma \frac{dx}{dt}=-\frac{\partial V(x)}{\partial x} + \xi(t) ,
\end{equation}
where the Gaussian  random force $\xi(t)$ satisfies  $\langle \xi(t) \rangle =0$ 
and  $\langle \xi(t) \xi(t^{'}) \rangle =2 \gamma T \delta(t-t^{'})$. 
Here $\langle ... \rangle$ denotes thermal average at temperature T(written for 
$k_{B}T$,~  where $k_{B}$ is the Boltzmann constant).~$\gamma$ is the frictional 
drag coefficient.~  The coresponding Fokker-Planck equation~~(in dimensionless units) is written as [3]
\begin{eqnarray}
\frac{\partial P(x,t)}{\partial t}  = {\bf{L}}_{FP}{P(x,t)} ,
\end{eqnarray}
where the Fokker-Planck operator ${\bf{L}}_{FP}$ is given by 
\begin{eqnarray}
{\bf{L}}_{FP}={\frac{\partial}{\partial x}} (V^{'}(x)+T \frac{\partial }
{\partial x}).
\end{eqnarray}
The prime over $V(x)$ denotes derivative with respect to {\it x}.~~After constructing the Hermitian operator $ {\bf{L}}=e^{\Phi/2}{\bf{L}}_{FP} 
e^{-\Phi/2}$,~ with the effective potential   $\Phi(x)=\frac{V(x)}{T}$,~~
the time indepedent Schr\"{o}dinger equation corresponding to the Fokker-Planck equation (2)  can be written as [3]
\begin{equation}
{\bf{L}}\psi_{n}(x)=[T\frac{\partial^{2}}{\partial x^{2}}-V_{S+}(x)]\psi_{n}(x)
\end{equation}
\begin{equation}
{\bf{L}}\psi_{n}(x)=-\lambda_{n} \psi_{n}(x)
\end{equation}
The Schr\"{o}dinger eigenfunction $\psi_{n}(x)=
e^{-\Phi(x)/2} \phi_{n}(x)$,~and the eigenvalues $\lambda_{n}(\geq 0)$ 
are given by
\begin{equation}  
{\bf{L}}_{FP} {\phi_{n}(x)}=-\lambda_{n} \phi_{n}(x)
\end{equation}
Here T substitutes for $\frac{\hbar^2}{2m}$ in quantum mechanical Schr\"{o}dinger equation.~ The Schr\"{o}dinger potential $V_{S+}(x)$ is obtained for the potential $V(x)$ as 
\begin{equation}
V_{S+}(x)=-\frac{V^{''}}{2}+\frac{{V^{'}}^{2}}{4T}
\end{equation}
For the potential $-V(x)$ the corresponding Schr\"{o}dinger potential is 
\begin{equation}
V_{S-}(x)=\frac{V^{''}}{2}+\frac{{V^{'}}^{2}}{4T}
\end{equation}
The potentials  $V_{S{\pm}}(x)=W^{2}(x) {\mp} \sqrt{T} W^{'}(x)$
constitute the supersymmetric partner potentials and $W(x)$ is referred to as the superpotential in supersymmetric quantum mechanics literature.~ The superpotential
$W(x)$   can also be expressed in terms of the ground state wavefunction of the Hermitian operator {\bf{L}} [4]. 
For periodic potentials ${\pm V(x)}$ in Fokker-Planck dynamics,~ it is known from  supersymmetric quantum mechanics that the eigenvalue spectra~ (with energy bands and band gaps) of  $ V_{S\pm}(x)$ of the corresponding Schr\"{o}dinger equation in quantum mechanics are strictly  isospectral and the eigenfunctions are directly related to each other being in one to one correspondence [4].~ The immediate implication of this observation is  that the  potentials ${\pm V(x)}$ will exhibit identical transport behavior in the long time asymptotic limit.~ It is expected that the average or expectation values of all the dynamical variables for $\pm V(x)$ will show similar behavior in the time asymptotic regime[4].~In fact for a given potential $V(x)$ in the presence of applied field F the analytic expressions for current,~J and diffussion coefficient,~D, are given by[7,8]
\begin{equation}
 \label{curr}
  J = 2 \pi \frac{1-\exp(-2 \pi F/T)}{\int^{x_{0}+2 \pi}_{x_{0}} \frac{dx}{2 \pi}
I_{\pm}(x)},\\
\end{equation}
and
\begin{equation}
  \label{quad}
  D = D_{0}\frac{\int_{x_{0}}^{x_{0}+2\pi} dx
    I_{\pm}(x)I_{+}(x)I_{-}(x)}{[\int_{x_{0}}^{x_{0}+2 \pi} dx I_{\pm}(x)]^{3}},
\end{equation}
where,
\label{I+-}
\begin{equation}
I_{+}(x) = \frac{1}{D_{0}} e^{\frac{V(x)-Fx}{T}}
\int_{x-2\pi}^{x} dy e^{\frac{-V(y)+Fy}{T}}, 
\end{equation}
\begin{equation}
I_{-}(x) = \frac{1}{D_{0}} e^{\frac{-V(x)+Fx}{T}}
\int_{x}^{x+2 \pi} dy e^{\frac{V(y)-Fy}{T}}.
\end{equation}
Here $2\pi$ is the length of the period and $D_{0}$ is the bare diffusion constant
($=kT/\gamma$).~ By tedious  algebraic manipulation~ (for example see the appendix of ref [7]) one can show that,~  for potentials  ${\pm V(x)}$,~ current  $J$ and diffusion coefficient $D$,~  respectively,~  satisfy the relations  
$J_{+}(F)=-J_{-}(-F)$  and  $D_{+}(F)=D_{-}(-F)$. ~The subscripts ${\pm}$  refer  to the potentials   ${\pm V(x)}$. The  result obtained in references [2]  follows as a special  case when $F=0$, ~ namely $D_{+}(F=0)=D_{-}(F=0)$.
\section{Results and Discussion} 
We have considered the potenitial $V(x)=A \cos{kx}+B \cos{2kx}$ in the Langevin equation in equation(1) in its dimensionless form.~We,~in the following,~ as has been done from  equation(2) onwards,~ measure energy in units of  $A$,~ wavenumber in units of $k$ and the friction coefficient in units of $\gamma$ and all other variables are given  in terms of combinations of $A$, $k$ and $\gamma$.~ Thus in these dimensionless units the potential reads as 
\begin{equation}
V(x)= \cos{x}+B \cos{2x},
\end{equation}
etc.~ Thus a particular choice of B will fix the relative values of the two potential barriers .
In a series of papers [2],~  the ratio of the diffusion constant to the bare diffusion constant~($D_{0}$) has been given in the entire range~(0,1) of the ratios of the potential barriers of $V(x)$ in the overdamped  as well as in the underdamped regime.~In these works they calculate the diffusion coefficients by solving the  Fokker-Planck equation (2) and (3) by the method of matrix continued fraction[3]. ~As mentioned in the previous sections their results  in the high friction limit,~ can be understood from supersymmetric considerations as well as obtained  from analytic expressions.~~We calculate in this work the diffusion constant $D$ and current $J$ in the presence of a constant force F.~ We calculate these quantities also for an asymmetric potential profile [fig.2]
\begin{equation}
V(x)= \cos{x}+ \cos{2x}+0.25 \sin{3x}.  
\end{equation}
In  the following  the nature of currents $J$ and the diffusion constant $D$  calculated  using  expressions (9) and (10) respectively are presented.

Fig.3 shows the variation of   current as a function of the constant applied force
$F$ at temperature $T=0.1$ for the symmetric potential $V(x)=\cos{x}+\cos{2x}$.~ As expected,~  for low values of $\vert F \vert << F_{c}$ (where the largest barrier to motion just disappears),~ the current effectively remains zero as the largest barrier height is much larger than  $T=0.1$ (at $F=0$  the largest barrier height is 3.0).~~It may be noted that at temperature $T=0$,~ there will be no current up to the critical field $F_{c}(\approx 2.74)$ as the particle will be in a locked state.~Beyond $F_{c}$ the particle being in a running state leads to a finite current[3].~At finite temperature current begins to flow even before $F_{c}$ due to thermal activation.~J approaches its linear regime (mobility 
$\mu=1$) soon after around $F$ equals ~$3$.~ From the figure it is clear that $J_{\pm}(F)=-J_{\pm}(-F)$ as also  $J_{+}(F)=J_{-}(F)$,~where the subscripts (${\pm}$) on $J$ corresponds to the potentials $\pm V(x)$.~Our results also corroborate the earlier obtained result that the diffusion constant acquires giant enhancement[7,8] at the critical field $F=F_{c}$~~[inset of fig.3].~The giant enhancement in $D$ near $F_{c}$ is expected as a fall out of the instability between the locked state ($F<F_{c}$) and running state[7,8].~We also observe that both the potentials ${\pm}V(x)$ exhibit identical diffussive behavior $D_{+}(F)=D_{-}(F)$ for this symmetrical potential.~~In fig.4,~ we plot $J_{\pm}$  and $D_{\pm}$  for the periodic potential ${\pm}V(x)$,~ with  $V(x)= \cos{x}+ \cos{2x}+0.25 \sin{3x}$,~  for the same temperature $T=0.1$.~Note that the crictical field now have different values ($F_{c1} \approx -3.19$ and $F_{c2} \approx 2.12$) for different applied field directions.  With the asymmetric potential  we see that the relation $J_{+}(+F)=-J_{-}(-F)$  is satisfied as mentioned in the introduction.~ $J(F)$   shows very similar behavior  to what is shown in fig.3.~The figure in the inset shows  $D_{+}(F)=D_{-}(-F)$.~For both the potentials the diffusion coefficient shows the resonant enhancement close to the corresponding critical fields $F_{c1}$ and $F_{c2}$.~And the diffusion coefficient does  show the symmetry with respect to $\pm F$,~ for either of the potentials $\pm V(x)$:  $D(F) \neq D(-F)$,~ however,~$D_{+}(F)=D_{-}(-F)$.

So far  we have considered the case for particle subjected to constant force.~We now study particle subjected to ac forcing where particle probability density  does not evolve to a steady state(rather it evolves to a time periodic state). It may be noted that supersymmetric arguments do not apply when time dependent forcings  are included. However,~even in this case we show numerically that diffusion constant shows the same behaviour for potentials $\pm V(x)$.~~In  Figs.~5 and 6 we plot the diffusion coefficient,~as a result of an ac drive 
$F=A \cos(\omega t)$  ,~ as a function of $A$  at temperature $T=0.25$.~~This is shown by solving the corresponding Langevin equation using Huen's method[9].~ The diffusion constant $D$  is  obtained in the long time asymptotic limit as 
\begin{equation}
D=\lim_{t \rightarrow \infty} {\frac{1}{2t}} [\langle x^2(t) \rangle -{\langle x(t) \rangle}^2]
\end{equation}
Where $\langle ... \rangle$ denotes the ensemble averaging over a large number of trajectories.~We discard the initial transients ($t_{0}=500 \tau$) and  allowed the system to evolve for $t=25000 \tau$. The time step $\tau$ is taken equal to 0.01 and 3500 ensembles are used to evaluate the averages (15).

In fig.5,~we give the plot for the  symmetric potential $V(x)=\cos{x} +\cos{2x}$ for two frequencies  $\omega=1.0$~ and $\omega=10.0$(inset).~It is very clear from the figure that $D$ is same for both the potentials $\pm V(x)$ within our numerical accuracy.~Also,~ for $\omega=1.0$,~ $D$ shows a resonant behaviour at $3.0<A<4.0$. It should again be noted that the largest barrier height shown by the potential $V(x)$  equals $3.0$.~That is, ~ ~peaking behavior is exhibited close to the extreme forcing equal to the critical static field $F_{c}$.~This is where the particle has maximum likelyhood of diffusing forward as well as in the backward direction thereby enhancing the diffusion coefficient.~For amplitudes A larger than about 3.0,~ the diffusion coefficients  are larger than their bare diffusion coefficient values[7,8].~Thus this peak could be attributed to optimised enhancement of the escape rate by modulation for a given noise strength[9,10,11].~It will be interesting  to see how the peaking behaviour of $D$ gets affected by the variation of temperature and frequency.~For high frequency drive ($\omega =10$,~for example),~ however, the peak disappears completely(Inset,~fig.5).~ Not only the peak disappears,~ the overall diffusivity itself gets reduced by about two orders of magnitude indicating that the particle excursion gets confined to a single well with only occasional jump to the right and left of the barriers,~exhibiting same behaviour  for both $\pm V(x)$. In fig.6,~ we plot the diffusion coefficient for the asymmetric potential
$V(x)=\cos{x}+\cos{2x}+0.25 \sin{3x}$ as a function of amplitude $A$ of the ac forcing $A\cos(\omega t)$ with  $\omega=1.0$  at temperature  $T=0.25$.~In this 
case too $D$ shows similar behavior to the symmetric potential case,~ fig.5.  Remarkably,~ also for the asymmetric potential $D_{+}(A)=D_{-}(A)$.
\section{Conclusion} 
In conclusion,~  we have shown that the current and diffusion coefficients,~ in static potentials $\pm V(x)$ in the presence of constant force,~are related by symmetry relations $J_{+}(F)=-J_{-}(-F)$ and $D_{+}(F)=D_{-}(-F)$,~respectively.~
When the Fokker-Planck equation of motion is mapped on to the Schr\"{o}dinger 
equation these potentials lead to supersymmetric Schr\"{o}dinger partner potentials, whose eigenspectra are isospectral.~Also,~ eigenfunctions are in one to one correspondence.~Consequently,~in the large time asymptotic limit these differing potentials  give the stated behaviour of the  transport coefficients.~We have also shown,~numerically,~that in the presence of ac drive the  diffusion coefficients  exhibit same behaviour for $\pm V(x)$.~~This is surprising because supersymmetry arguments do not apply in the presence of time dependent fields.~This~ ,~therefore,~ requires further investigation.
\section{Acknowledgement}
A.M.J. and M.C.M. thank BRNS,~DAE,~Mumbai for partial financial support.
A.M.J.  thanks  Professor  A.~Khare for  valuable discussion.~M.C.M. gratefully acknowledges Institute of Physics,~Bhubaneswar for hospitality.
\newpage
\Large{References:}

\newpage

\begin{figure}[htp!]
\begin{center}
\input{epsf}
\includegraphics [width=3.5in,height=4.5in] {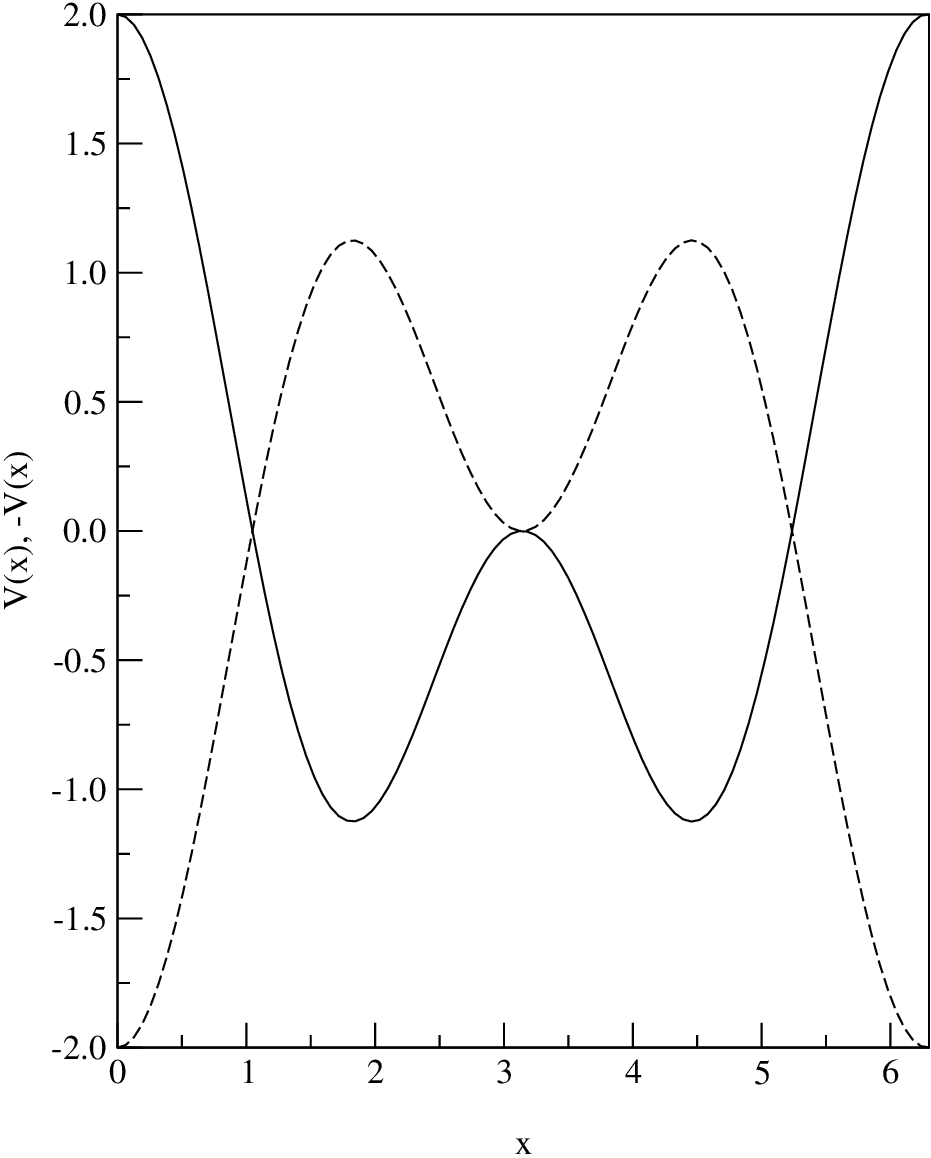}
\caption{The potentials $V(x)=\cos{x}+\cos{2x}$ (solid line)  and $-V(x)$ (dashed
line) are plotted in dimensionless units for comparision.} 
\label{fig-symmetric potential}
 \end{center}
 \end{figure}
\newpage
\begin{figure}[htp!]
\begin{center}
\input{epsf}
\includegraphics [width=3.5in,height=4.5in] {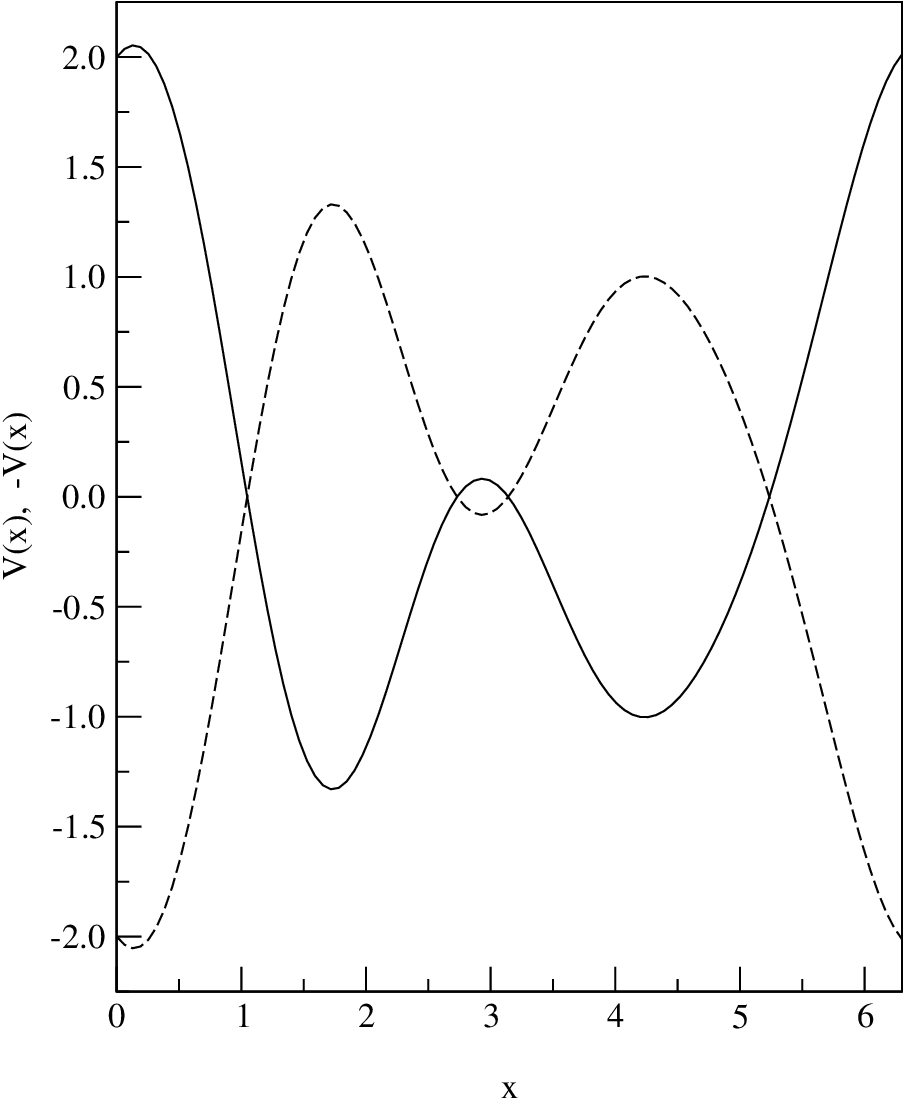}
\caption{The potentials $V(x)=\cos{x}+\cos{2x}+0.25\sin{3x}$(solid line)  
and $-V(x)$ (dashed line) are plotted in dimensionless units for comparision.} 
\label{fig-symmetric potential}
 \end{center}
 \end{figure}
\newpage
\begin{figure}[htp!]
\begin{center}
\input{epsf}
\includegraphics [width=4.0in,height=4.5in] {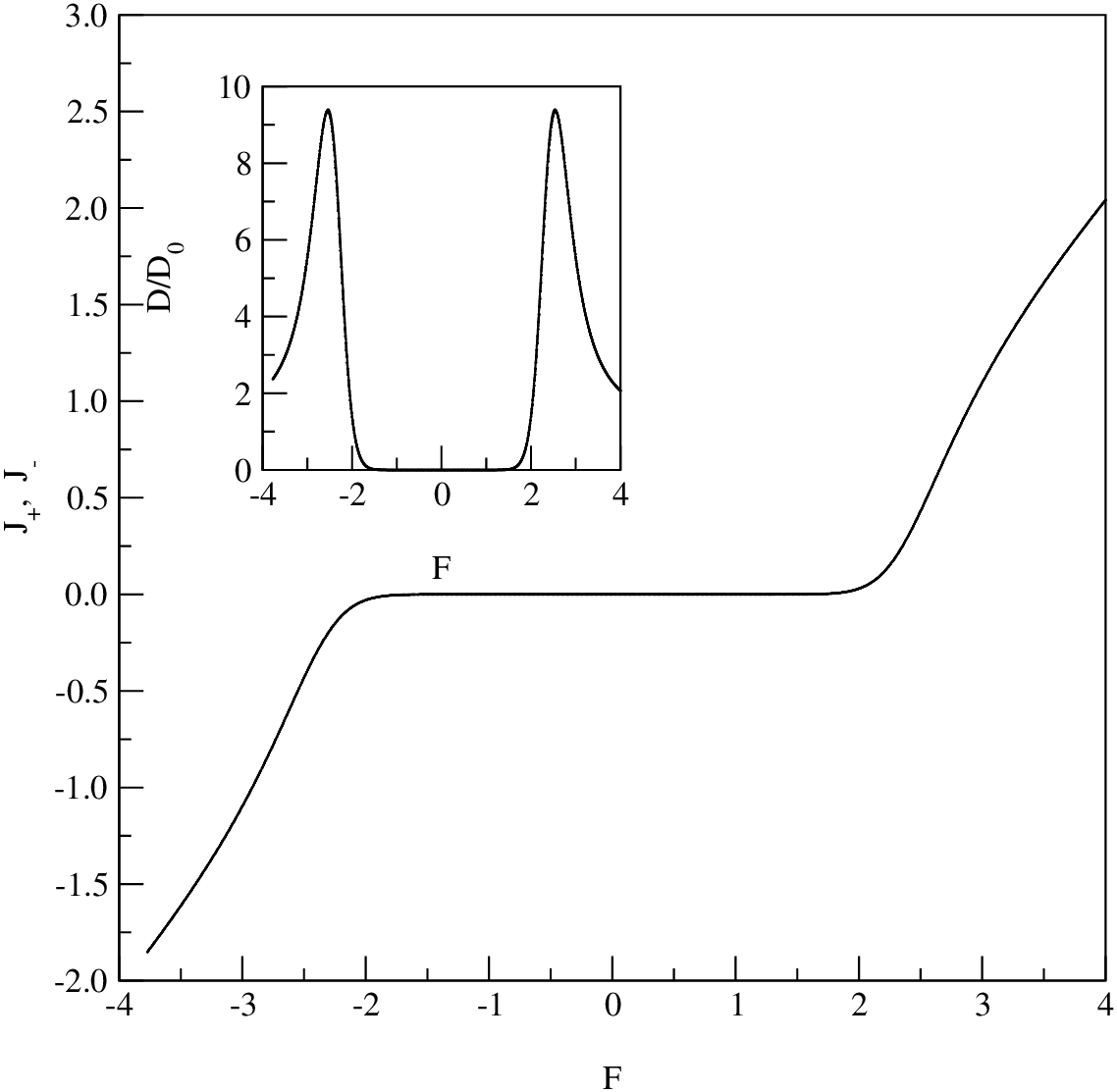}
\caption{The currents $J_{\pm}$   and   (inset) $D_{\pm}$  are plotted as a 
function of constant applied force $F$ for symmetric potentials.~~The subscripts 
 $\pm$  refer to the potentials $\pm V(x)$  and at temperature $T=0.1$.~~
Results for $\pm V(x)$ are identical. } 
\label{fig-symmetric potential}
 \end{center}
 \end{figure}
\newpage

\begin{figure}[htp!]
\begin{center}
\input{epsf}
\includegraphics [width=4.0in,height=4.5in] {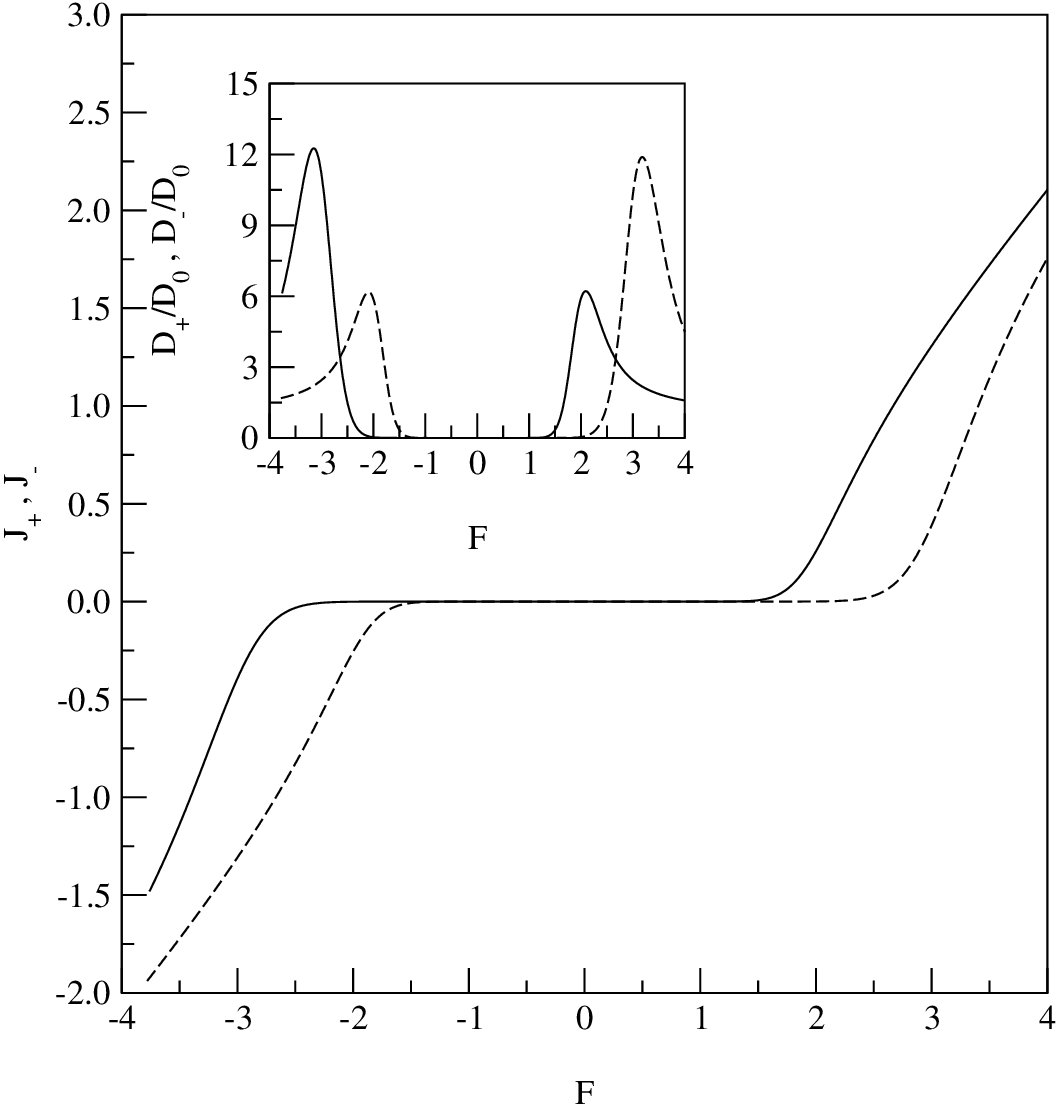}
\caption{The currents $J_{\pm}$   and   (inset) $D_{\pm}$  are plotted as a
function of constant applied force $F$   for asymmetric potentials.~~The subscripts $\pm$ refers to the potentials $\pm V(x)$  and at temperature 
$T=0.1$.~~Continuous lines correspond to $+V(x)$.  } 
\label{fig-symmetric potential}
 \end{center}
 \end{figure}
\newpage

\begin{figure}[htp!]
\begin{center}
\input{epsf}
\includegraphics [width=4.0in,height=4.5in] {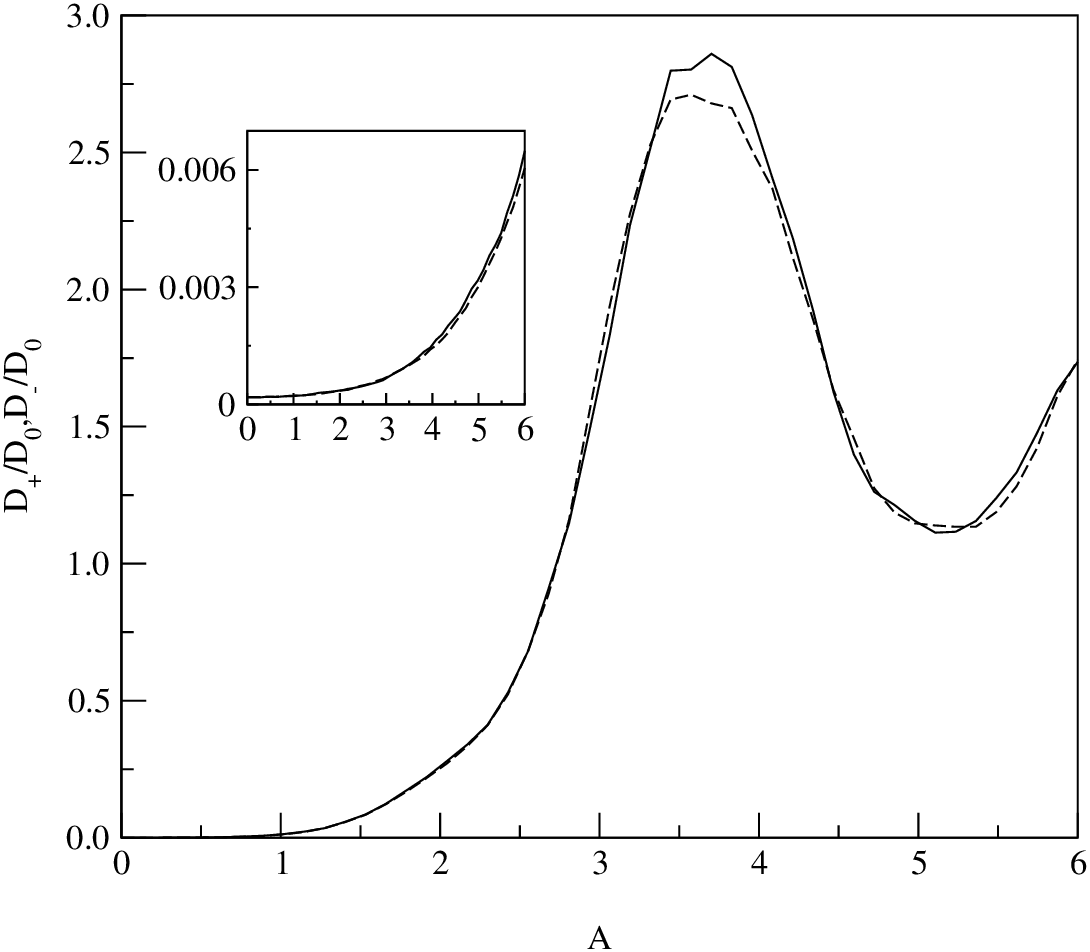}
\caption{The diffusion constants $D_{\pm}$ are plotted as a function of the 
amplitude A of ac force for the symmetric potential  $\pm V(x)$ at frequencies
 $\omega=1.0$ and (inset) $\omega=10.0$ and at temperature $T=0.25$. The 
continuous line corresponds to $D_{+}$.} 
\label{fig-symmetric potential}
 \end{center}
 \end{figure}

\newpage

\begin{figure}[htp!]
\begin{center}
\input{epsf}
\includegraphics [width=4.0in,height=4.5in] {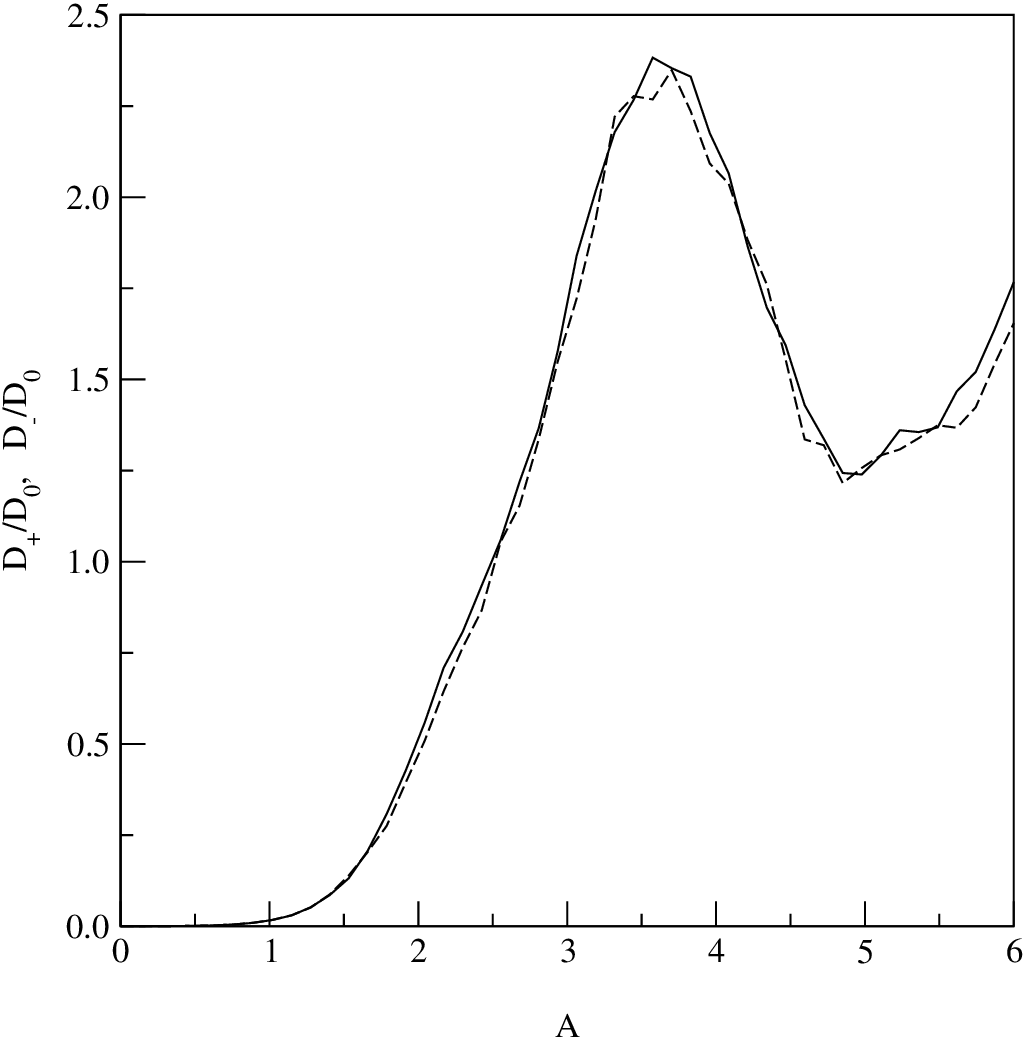}
\caption{ The variation of $D_{\pm}$ as a function of amplitude $A$ of a ac 
drive for the asymmetric potential $\pm V(x)$,~~ (eq.(14))~ at temperature $T=0.25$
and high frequency  $\omega=1.0$.~ Continuous line for $+V(x)$.} 
\label{fig-asymmetric potential}
\end{center}
\end{figure}
\end{document}